\documentclass[12pt]{article}

\usepackage{indentfirst} 
\usepackage{amsfonts,amssymb,amsmath}
\usepackage{bm}          
\usepackage{cite}
\usepackage{url}
\usepackage{enumitem}
\usepackage[dvipsnames]{xcolor}
\usepackage{blkarray}
\usepackage{multirow}

\usepackage{MnSymbol}
\usepackage{hyperref}
\usepackage{tensor}
\usepackage{arcs}
\usepackage{tikz-cd}
\usepackage{todonotes}
\usepackage[titletoc,title]{appendix}
\usepackage{tikz}
\usetikzlibrary{decorations.pathreplacing,positioning,matrix}
\usetikzlibrary{arrows}

\AtBeginDocument{}%

\newcommand{\corurl}{red}
\newcommand{\corcite}{ForestGreen}
\newcommand{\corlink}{blue}
\newcommand{\dd}{\mathrm{d\!\raisebox{-0.05pt}{\scalebox{1}[1.021]{l}}}}
\newcommand{\w}{\wedge\hspace*{-6.77pt}{\raisebox{0.08pt}{\scalebox{.75}[.75]{$\wedge$}}}}
\newcommand{\ww}{\wedge\hspace*{-9.10pt}{\scalebox{.75}[.75]{$\wedge$\,\,\,}}}

\newcommand{\bd}{\bm{\mathrm{d}}}

\newcommand{\Aa}{\bm{\mathrm{A}}}

\newcommand{\wed}{\wedge\hspace*{-7.5pt}.\hspace*{3pt}}
\newcommand{\wedgmond}{
\mathrel{
    \makebox[4pt][c]{
      \begin{tikzpicture}[baseline=-0.5ex, scale=0.7]
        \node at (0,0) {\large$\wedge$};
        \node at (0,-0.15) {\tiny$\diamond$};
      \end{tikzpicture}
    }
  }
}

\hypersetup{linktocpage,colorlinks,urlcolor=\corurl,citecolor=\corcite,linkcolor=\corlink,
pdftitle={Carrollian limit Husain-Kucha\v{r} model},
pdfauthor={J.F. Barbero, J. Margalef-Bentabol, A. Vicente-Cano, E.J.S. Villase\~nor}}

\numberwithin{equation}{section}  

\def\QED{{\boldmath$\rule{0.5em}{0.5em}$}}                                
\def\markatright#1{\leavevmode\unskip\nobreak\quad\hspace*{\fill}{#1}}    
\def\qed{\markatright{\QED}}                                              

\usepackage{titling}
\usepackage{authblk}
\title{Husain-Kucha\v{r} model as the Carrollian limit of the Holst term}

\author[1]{J. Fernando Barbero G.\footnote{Corresponding author.} }
\author[2,3]{Juan Margalef-Bentabol\,}
\author[4]{Aitor Vicente-Cano\,}
\author[3,5]{Eduardo J.S. Villase\~nor\,}
\affil[1]{\href{https://ror.org/05rtchs68}{Instituto de Estructura de la Materia}, IEM-CSIC, Serrano 123, 28006 Madrid, Spain}
\affil[2]{\href{https://ror.org/0161xgx34}{University of Montreal}, Department of Mathematics and Statistics,  CP 6128 Succ. Centre-Ville Montréal, QC H3C3J7, Canada}
\affil[3]{Grupo de Teor\'{\i}as de Campos y F\'{\i}sica Estad\'{\i}stica. Instituto Gregorio Mill\'an (UC3M). Unidad Asociada al Instituto de Estructura de la Materia, CSIC}
\affil[4]{Departament de F\'{i}sica Qu\`antica i Astrof\'{i}sica, \href{https://ror.org/044fgj614}{Institut de Ci\`encies del Cosmos}, Universitat de Barcelona, Mart\'{i} i Franqu\`es 1, 08028 Barcelona, Spain}
\affil[5]{\href{https://ror.org/03ths8210}{Universidad Carlos III de Madrid}, Departamento de Matemáticas, Avenida de la Universidad, 30
 (edificio Sabatini), 28911, Legan\'es (Madrid), España}
\date{}                     
\begin{document}
	
\maketitle
\date{September 30, 2024}

\vspace*{-10ex}

\begin{abstract}
We show how the Husain-Kucha\v{r} model can be understood as a Carrollian limit of the Holst term in the context of background-independent field theories described in terms of coframes and spin connections. We also discuss the footprint of the Carrollian
symmetry in the Hamiltonian formulation of the Husain-Kucha\v{r} action.
\end{abstract}

\tableofcontents


%
%
	
\section{Introduction}\label{sec_Carrol_Introduction}

The Carrollian and Galilean limits of the Poincaré group \cite{BL,Levy-Leblond,sengupta} and many relativistic field theories have been studied for a long time (see, for example, \cite{D1,D2,Bergshoeff,Concha,Correa,tadros2024,hansen2022carroll}). They are natural simplifications of relativistic models valid in different contexts. On one hand, the Galilean limit $c\rightarrow\infty$ is useful whenever the relative velocities of the relevant physical objects are small. This is the limit of everyday life and also the one in which non-relativistic quantum mechanics is framed. On the other extreme is the Carrollian limit $c\rightarrow0$, in which the light cones collapse to lines and propagation in space-time is controlled by congruences of curves tangent to them.

An interesting family of background-independent field theories related to general relativity is comprised by the Husain-Kucha\v{r} (HK) model \cite{Husain:1990vz} and some close relatives \cite{Husain:1992qx,BarberoG:1997nrd,Husain:2023fus}. The main reason why the HK model has received significant attention in the past, especially within the context of Loop Quantum Gravity (LQG), is because its Hamiltonian description is very close to the Ashtekar formulation for General Relativity (GR), but significantly simpler. In fact, the most relevant difference between the two Hamiltonian descriptions is the absence of the scalar constraint in the HK model. This means that one of the main difficulties in the study of the classical dynamics and the canonical quantization of GR is absent, while some crucial features --in particular diff-invariance or background independence-- are retained.


The purpose of this paper is to show how the HK model can be understood as a Carrollian limit in the context of background-independent field theories described in terms of coframes and spin connections. Regarding this issue, an important point must be made right from the start. In the early literature on the subject \cite{Husain:1993he,Rovelli:1992vv}, it was suggested that the HK model could be derived in the usual metric variables by a limiting procedure in which the speed of light $c$ is taken to zero. This result appears to be natural (basically, one just has to find an argument proving that the scalar constraint disappears in this limit), but as we will show, in the context of the 1-form variables in terms of which the original HK model was written, it actually comes from the Carrollian limit of an action consisting \textit{only} \cite{Figueroa-OFarrill:2022mcy} on the so-called Holst term \cite{Holst:1995pc}. (The Hamiltonian analysis of the action defined solely by the Holst term, before taking the Carrollian limit, can be found in \cite{Liu:2009em}. The approach presented there may possibly be used as an alternative starting point to part of the present work). It is interesting to point out that, to our knowledge, there are no actions that lead to a version of the HK model in terms of metric variables (i.e. to a Hamiltonian formulation where the 3-metric is constrained to satisfy the vector constraint of the ADM formulation but there is no scalar constraint) although it is conceivable that it can be obtained by considering the parity violating term introduced in \cite{Hojman:1980kv}.

This paper provides preliminary information necessary to interpret the dynamics of general Carrollian gravity models, such as the ones presented in \cite{Figueroa-OFarrill:2022mcy}. A systematic derivation and a complete understanding of the Hamiltonian description of these models is difficult and requires a gradual approach for which this work is a first step. As we will discuss in the following, the geometric structure of the collapsed Carrollian light cones in the HK model can be found both from the field equations and the Hamiltonian dynamics. We expect that this kind of structure will also be present in general Carrollian models but will be harder to spot, hence it is important to understand the one instance in which we have sufficient control on this issue. We also expect that the simplicity of the dynamics of the HK model which, in retrospective, is a consequence of the collapse of the light cones, will be a generic feature of all the Carrollian gravitational theories. A possible application of these theories, as suggested in \cite{Sengupta:2022rbd}, is in the context the BKL conjecture.

A final point that we will discuss has to do with the interpretation of the Carrollian symmetry and its footprint in the Hamiltonian formulation. As has been discussed in the literature \cite{Figueroa-OFarrill:2022mcy}, the standard HK action can be obtained as an invariant term by a systematic procedure devised to find all the Carroll-symmetric actions (which can be written in terms of differential forms). The imprint of the Carrollian symmetry must be of a different type because (i) the procedure described in \cite{Figueroa-OFarrill:2022mcy} respects the invariance under diffeomorphisms and (ii) the Carrollian symmetry cannot be directly read from the Hamiltonian formulation for the HK action. We argue in the paper that this has to do with the triviality of the dynamics in the concrete sense that we explain later.

The structure of the paper is the following. After this introduction, we introduce in Section \ref{sec_geometric arena} the relevant geometric framework that combines elements of both \cite{Figueroa-OFarrill:2022mcy} and \cite{Bergshoeff:2017btm}. We will show in Section \ref{sec_Carrol_Holst} how the HK action appears after a suitable limiting procedure starting from the Holst action.  In Section \ref{sec_HK}, we review the relevant facts about the HK model that we will use in the paper. In Section \ref{sec_light_cones}, we will discuss the issues mentioned in this introduction. The paper ends with some conclusions and comments in Section \ref{sec_conclusions} and an Appendix \ref{app_contractions} where we give some mathematical details about group contractions.

\section{Geometric arena}\label{sec_geometric arena}

\subsection{The Poincar\'e algebra}\label{subsec_Poincare}

Consider the ${(-}{+}{+}{+)}$ (internal) Minkowski metric $\eta_{\alpha\beta}$ ($\alpha=0,1,2,3$). Latin indices are raised and lowered with the Euclidean metric $\delta_{ij}$ and $\delta^{ij}$ and Lorentz indices with $\eta_{\alpha\beta}$ and $\eta^{\alpha\beta}$. The 3-dimensional Levi-Civita symbol is $\varepsilon_{ijk}$ and the Minkowskian one $\varepsilon_{\alpha\beta\gamma\delta}$ ($\varepsilon_{0123}=+1$). We will work with fields defined on a four-dimensional manifold $\mathcal{M}$. These fields will be denoted with boldface letters. We do the same with the exterior and covariant differentials (however, we will use standard, non-boldface letters in the 3-dimensional manifold used in the Hamiltonian analysis).

We start by writing the commutators of the generators of the $\mathfrak{so}(1,3)$ Lie algebra
\begin{equation}\label{Lorentz3plus1}
[L_{\alpha\beta},L_{\gamma\delta}]=\eta_{\alpha\delta}L_{\gamma\beta}-\eta_{\alpha\gamma}L_{\delta\beta}-\eta_{\beta\delta}L_{\gamma\alpha}+\eta_{\beta\gamma}L_{\delta\alpha}\,,
\end{equation}
where we employ the usual $\mathfrak{so}(1,3)$ basis spanned by $L_{\alpha\beta}$ with $0\leq\alpha<\beta\leq 3$ and, for convenience, introduce $L_{\alpha\alpha}:=0$ for all $\alpha$ and $L_{\beta\alpha}:=-L_{\alpha\beta}$ when $\beta> \alpha$. The Poincaré Lie algebra in $1+3$ dimensions is spanned by a basis consisting of the Lorentz generators $L_{\alpha\beta}$ plus the generators $P_{\alpha}$ of the translations in Minkowski spacetime. The brackets of these are
\begin{align}
[L_{\alpha\beta},L_{\gamma\delta}]&=\eta_{\alpha\delta}L_{\gamma\beta}-\eta_{\alpha\gamma}L_{\delta\beta}-\eta_{\beta\delta}L_{\gamma\alpha}+\eta_{\beta\gamma}L_{\delta\alpha}\,,\label{LL}\\
[L_{\alpha\beta}\,,\,P_\gamma]&=\eta_{\gamma\alpha}P_\beta-\eta_{\gamma\beta}P_\alpha \,,\label{LP}\\
[\,P_{\,\alpha}\,\,,P_{\,\beta}]&=0\,.\label{PP}
\end{align}

In the following, we will separate boosts from rotations and write
\begin{equation}\label{boosts_rotations}
K_i:=L_{0i}\,,\quad \varepsilon_{ijk}J^k:=L_{ij}\,\, \Leftrightarrow J^i:=\frac{1}{2}\varepsilon^{ijk}L_{jk}\,,\quad j=1,2,3\,,
\end{equation}
The purely ``temporal terms'' $\{P_0,K_i\}$ generate the subspace that, following the notation of Appendix  \ref{app_contractions}, we denote $\mathfrak{m}$. Meanwhile, the ``spatial elements'' $\{P_i,J_i\}$ generate the subspace denoted $\mathfrak{h}$. The brackets of the basis elements $\{P_0,K_i,P_i,J_i\}$ become
\[\begin{blockarray}{lllll}
     \BAmulticolumn{2}{l}{\overbrace{\hspace{13em}}^{\hspace{0em}\scalebox{1}{$[\cdot{},\mathfrak m]$}}} &
     \BAmulticolumn{2}{l}{\overbrace{\hspace{16.5em}}^{\hspace{0em}\scalebox{1}{$[\cdot{},\mathfrak h]$}}}  \\
    \begin{block}{llll\}c}
         [P_0,P_0]=0 \ \ \mbox{}& [P_0,K_j]=P_j & [P_0,P_j]=0&[P_0,J_j]=0  & \multirow{2}{*}{$[\mathfrak m,\cdot{}]$}\\
        & [K_i,K_j]=-\varepsilon_{ijk}J^k  \ \ \mbox{}&[K_i,P_j]=-\delta_{ij}P_0  & [K_i,J_j]=\varepsilon_{ijk}K^k &  \\
    \end{block}
    \begin{block}{llll\}c}
        && [P_i,P_j]=0\ \ \mbox{} & [P_i,J_j]=\varepsilon_{ijk}P^k &\multirow{2}{*}{$[\mathfrak h,\cdot{}]$} \\
          &  &   & [J_i,J_j]=\varepsilon_{ijk}J^k\ \ \mbox{} &  \\
    \end{block}
\end{blockarray}\]

Notice that $[\mathfrak{m},\mathfrak{m}]\subset \mathfrak{h}$, $[\mathfrak{h},\mathfrak{h}]\subset \mathfrak{h}$, and $[\mathfrak{h},\mathfrak{m}]\subset \mathfrak{m}$.

\subsection{The Carrollian limit of the algebra}\label{subsec_Carrol_Holst}
In order to discuss the Carrollian limit of the Holst action, we start by rescaling the generators of $\mathfrak{iso}(1,3)$ as in \cite{Levy-Leblond}. Let us then consider the change of basis given by $H:=cP_0$ and $G_i:=cK_i$, where we can think of the parameter $c\neq0$ as the speed of light that we will later take to zero. In terms of this new basis $\{H,J_i,P_i,G_i\}$, the Lie algebra of $\mathfrak{iso}(1,3)$ takes the form
\[\begin{blockarray}{llll}
         [H,H]=0 \ \ \mbox{}& [H,G_j]=c^2P_j & [H,P_j]=0& [H,J_j]=0   \\
        & [G_i,G_j]=-c^2\varepsilon_{ijk}J^k   \ \ \mbox{}& [G_i,P_j]=-\delta_{ij}H & [G_i,J_j]=\varepsilon_{ijk}G^k  \\
        && [P_i,P_j]=0\ \ \mbox{} &[P_i,J_j]=\varepsilon_{ijk}P^k  \\
          &  &   & [J_i,J_j]=\varepsilon_{ijk}J^k \\
\end{blockarray}\]
By contracting the algebra with $c\to 0$ (see Appendix \ref{app_contractions} for further details), we get the $1+3$ dimensional Carroll algebra \cite{Levy-Leblond}
\[\begin{blockarray}{llll}
         [H,H]=0\qquad \ \ \mbox{}& [H,G_j]=0\qquad & [H,P_j]=0& [H,J_j]=0   \\
        & [G_i,G_j]=0\qquad   \ \ \mbox{}& [G_i,P_j]=-\delta_{ij}H & [G_i,J_j]=\varepsilon_{ijk}G^k  \\
        && [P_i,P_j]=0\ \ \mbox{} &[P_i,J_j]=\varepsilon_{ijk}P^k  \\
          &  &   & [J_i,J_j]=\varepsilon_{ijk}J^k \\
\end{blockarray}\]
Notice that with this new bracket, we still have $[\mathfrak{m},\mathfrak{m}]\subset \mathfrak{h}$, $[\mathfrak{h},\mathfrak{h}]\subset \mathfrak{h}$, and $[\mathfrak{h},\mathfrak{m}]\subset \mathfrak{m}$.
\subsection{The Carrollian split of the geometric objects}{\label{defcarr}}
To write the action, we consider the $\mathfrak{iso}(1,3)$ invariant form in the algebra, which we will denote as $\langle\cdot,\cdot\rangle$, whose only non-zero terms are 
\begin{equation}
\langle L_{\alpha\beta}\,,L_{\gamma\delta}\rangle=\eta_{\alpha\gamma}\eta_{\beta\delta}-\eta_{\beta\gamma}\eta_{\alpha\delta}\,.
\end{equation}
It is $\mathrm{Ad}$-invariant and degenerate (because $\mathfrak{iso}(1,3)$ is not semi-simple). In the $\{H,P_i,G_i,J_i\}$ basis, the only non-zero products are
\begin{equation}\label{eq: products G and J}
    \langle G_i,G_j\rangle=-c^2\delta_{ij}\,,\quad    \langle J_i,J_j\rangle=\delta_{ij}\,.
\end{equation}
Let ${\bm{\mathrm{C}}}$ be an $\mathfrak{iso}(1,3)$-valued connection 1-form whose expansion in the $\{L_{\alpha\beta},P_\alpha\}$  basis is
\[
{\bm{\mathrm{C}}}={\bm{\mathrm{e}}}^\alpha P_\alpha+\frac{1}{2}{\bm{\mathrm{W}}}^{\alpha\beta}L_{\alpha\beta}=:{\bm{\mathrm{e}}}+{\bm{\mathrm{W}}}\,,
\]
where ${\bm{\mathrm{W}}}^{\alpha\beta}=-{\bm{\mathrm{W}}}^{\beta\alpha}$. Its associated $\mathfrak{iso}(1,3)$-valued curvature 2-form is
\begin{equation}\label{F}
{\bm{\mathrm{F}}}_{\mathrm{C}}:=\bm{\mathrm{d}}{\bm{\mathrm{C}}}+\frac{1}{2}[{\bm{\mathrm{C}}}\!\wed {\bm{\mathrm{C}}}]\,,
\end{equation}
where we use the following notation: if ${\bm{\mathrm{A}}}_p\in\Omega^p(\mathcal{M},\mathfrak{g})$ and ${\bm{\mathrm{A}}}_q\in\Omega^q(\mathcal{M},\mathfrak{g})$ are expanded on a basis $(X_i)$ of a Lie algebra $\mathfrak{g}$ as ${\bm{\mathrm{A}}}_p={\bm{\mathrm{A}}}_p^iX_i$ and ${\bm{\mathrm{A}}}_q={\bm{\mathrm{A}}}_q^iX_i$ with ${\bm{\mathrm{A}}}_p^i\in\Omega^p(\mathcal{M})$ and ${\bm{\mathrm{A}}}_q^i\in\Omega^q(\mathcal{M})$, we define
\begin{align*}
&[{\bm{\mathrm{A}}}_p\wed {\bm{\mathrm{A}}}_q]:={\bm{\mathrm{A}}}_p^i\wedge {\bm{\mathrm{A}}}_q^j[X_i,X_j]\in\Omega^{p+q}(\mathcal{M},\mathfrak{g})\,,\\
&\langle{\bm{\mathrm{A}}}_p\wed {\bm{\mathrm{A}}}_q\rangle:={\bm{\mathrm{A}}}_p^i\wedge {\bm{\mathrm{A}}}_q^j\langle  X_i,X_j\rangle\in\Omega^{p+q}(\mathcal{M})\,.
\end{align*}
Writing the curvature in the $\{L_{\alpha\beta},P_\alpha\}$ basis, we obtain
\[{\bm{\mathrm{F}}}_{\mathrm{C}}=({\bm{\mathrm{D}}}_{\mathrm{W}}{\bm{\mathrm{e}}})^\alpha P_\alpha+\frac{1}{2}{\bm{\mathrm{F}}}_{\mathrm{W}}^{\alpha\beta}L_{\alpha\beta}=:{\bm{\mathrm{D}}}_{\mathrm{W}}{\bm{\mathrm{e}}}+{\bm{\mathrm{F}}}_{\mathrm{W}}\,\qquad\begin{array}{l}
({\bm{\mathrm{D}}}_{\mathrm{W}}{\bm{\mathrm{e}}})^\alpha:=\bm{\mathrm{d}}{\bm{\mathrm{e}}}^\alpha-{\bm{\mathrm{W}}}^{\alpha\beta}\!\!\wedge {\bm{\mathrm{e}}}_\beta\,,\\[1.5ex]{\bm{\mathrm{F}}}_{\mathrm{W}}^{\alpha\beta}:=\bm{\mathrm{d}}{\bm{\mathrm{W}}}^{\alpha\beta}-{\bm{\mathrm{W}}}^{\alpha}_{\phantom{\alpha}\gamma}\!\wedge\!{\bm{\mathrm{W}}}^{\gamma\beta}\,.
\end{array}\]
Notice that ${\bm{\mathrm{F}}}_{\mathrm{W}}=\!{\bm{\mathrm{d}}}{\bm{\mathrm{W}}}\!\!+\!\frac{1}{2}[{\bm{\mathrm{W}}}\!\wed {\bm{\mathrm{W}}}]$.

We can also expand ${\bm{\mathrm{e}}}$ and ${\bm{\mathrm{W}}}$ in the $\{H,P_i,G_i,J_i\}$ basis as follows:
\begin{align*}
    &{\bm{\mathrm{e}}}={\bm{\mathrm{\tau}}} H+{\bm{\mathrm{e}}}^iP_i\,,\qquad\text{with }{\bm{\mathrm{\tau}}}=\frac{1}{c}{\bm{\mathrm{e}}}^0\,,\\
    &{\bm{\mathrm{W}}}={\bm{\mathrm{\Omega}}}^iG_i+{\bm{\mathrm{A}}}^iJ_i\,,\qquad\text{with } {\bm{\mathrm{\Omega}}}^i=\frac{1}{c}{\bm{\mathrm{W}}}^{0i}\,,\quad {\bm{\mathrm{A}}}_k=\frac{1}{2}\varepsilon_{ijk}{\bm{\mathrm{W}}}^{ij}\,.
\end{align*}
This decomposition leads to:
\begin{align*}
&{\bm{\mathrm{C}}}={\bm{\mathrm{\tau}}} H+{\bm{\mathrm{e}}}^iP_i+{\bm{\mathrm{\Omega}}}^iG_i+{\bm{\mathrm{A}}}^iJ_i\,,\phantom{\bigg{|}}\\
&{\bm{\mathrm{D}}}_{\mathrm{W}}{\bm{\mathrm{e}}}=(\bm{\mathrm{d}}{\bm{\mathrm{\tau}}}-{\bm{\mathrm{\Omega}}}_i\wedge {\bm{\mathrm{e}}}^i)H+(\bm{\mathrm{d}}{\bm{\mathrm{e}}}^i+\varepsilon^i_{\phantom{i}jk}{\bm{\mathrm{A}}}^j\wedge {\bm{\mathrm{e}}}^k-c^2{\bm{\mathrm{\Omega}}}^i\wedge {\bm{\mathrm{\tau}}})P_i\,,\\
&{\bm{\mathrm{F}}}_{\mathrm{W}}=\big({\bm{\mathrm{d}}}{\bm{\mathrm{\Omega}}}^i+\varepsilon^i_{\phantom{i}jk}{\bm{\mathrm{A}}}^j\wedge {\bm{\mathrm{\Omega}}}^k\big)G_i+\Big({\bm{\mathrm{d}}}{\bm{\mathrm{A}}}^i+\frac{1}{2}\varepsilon^i_{\phantom{i}jk}{\bm{\mathrm{A}}}^j\wedge {\bm{\mathrm{A}}}^k-\frac{c^2}{2}\varepsilon^i_{\phantom{i}jk}{\bm{\mathrm{\Omega}}}^j\wedge {\bm{\mathrm{\Omega}}}^k\Big)J_i\,.
\end{align*}

\section{The Carrollian limit of the Holst action}\label{sec_Carrol_Holst}
In this section, we will analyze the Carrollian limit of the Holst action defined on a 4-dimensional space-time $\mathcal{M}$ without boundaries. To that end, we consider the antisymmetric bilinear map
\begin{align*}
&(\cdot\diamond\cdot):\mathfrak{iso}(1,3)\times\mathfrak{iso}(1,3)\rightarrow\mathfrak{iso}(1,3)\\
&(P_\alpha\diamond P_\beta)=\frac{1}{2}\beta_{\mathrm{P}}\,\varepsilon_{\alpha\beta}^{\phantom{\alpha\beta}\gamma\delta}L_{\gamma\delta}+\gamma L_{\alpha\beta}\,,\quad\beta_{\mathrm{P}}\,,\gamma\in\mathbb{R}\,,\\
&(P_\alpha\diamond L_{\beta\gamma})=0\,,\\
&(L_{\alpha\beta}\diamond L_{\gamma\delta})=0\phantom{\Big{|}}\,.
\end{align*}
It is straightforward to see that this map is equivariant under the action of $SO(1,3)$, i.e. if $R\in SO(3)$ then:
\begin{align*}
&\big(\mathrm{Ad}_RP_\alpha\diamond \mathrm{Ad}_RP_\beta\big)=\mathrm{Ad}_R(P_\alpha\diamond P_\beta)\,,\\
    &\big(\mathrm{Ad}_RP_\alpha\diamond \mathrm{Ad}_RL_{\beta\gamma}\big)=\mathrm{Ad}_R(P_\alpha\diamond L_{\beta\gamma})\,,\\
    &\big(\mathrm{Ad}_RL_{\alpha\beta}\diamond \mathrm{Ad}_RL_{\gamma\delta}\big)=\mathrm{Ad}_R(L_{\alpha\beta}\diamond L_{\gamma\delta})\,,
\end{align*}
where $\mathrm{Ad}_RP_\alpha=R P_{\alpha}R^{-1}$ and so on. In terms of the basis $\{H,P_i,G_i,J_i\}$, the only non-zero terms are:
\[
(H\diamond P_i)=c\beta_{\mathrm{P}}\, J_i+\gamma G_i\,,\quad (P_i\diamond P_j)=\varepsilon_{ijk}\left(\gamma J^k-\frac{\beta_{\mathrm{P}}}{c}G^k\right)\,.
\]

We finally consider the Holst Lagrangian
\begin{equation}\label{Lagrangian}
    \mathcal{L} =\langle ({\bm{\mathrm{e}}}\wedgmond {\bm{\mathrm{e}}})\wed {\bm{\mathrm{F}}}_{\mathrm{W}}\rangle\,,
\end{equation}
where
\begin{align*}
({\bm{\mathrm{e}}}\wedgmond  {\bm{\mathrm{e}}}):\!&={\bm{\mathrm{e}}}^\alpha\wedge {\bm{\mathrm{e}}}^\beta (P_\alpha\diamond P_\beta)\\
&=\Big(2\gamma{\bm{\mathrm{\tau}}}\wedge {\bm{\mathrm{e}}}^i-\frac{\beta_{\mathrm{P}}}{c}\varepsilon^i_{\phantom{i}jk}{\bm{\mathrm{e}}}^j\wedge {\bm{\mathrm{e}}}^k\Big)G_i+\big(2c\beta_{\mathrm{P}}{\bm{\mathrm{\tau}}}\wedge {\bm{\mathrm{e}}}^i+\gamma\varepsilon^i_{\phantom{i}jk}{\bm{\mathrm{e}}}^j\wedge {\bm{\mathrm{e}}}^k\big)J_i\,.
\end{align*}
Introducing this expression into \eqref{Lagrangian} and using \eqref{eq: products G and J}, one finds
\begin{align*}
\mathcal{L}=&-c^2\Big(2\gamma{\bm{\mathrm{\tau}}}\wedge {\bm{\mathrm{e}}}^i-\frac{\beta_{\mathrm{P}}}{c}\varepsilon^i_{\phantom{i}jk}{\bm{\mathrm{e}}}^j\wedge {\bm{\mathrm{e}}}^k\Big)\wedge({\bm{\mathrm{d}}}{\bm{\mathrm{\Omega}}}_i+\varepsilon_{i\ell m}{\bm{\mathrm{A}}}^\ell\wedge {\bm{\mathrm{\Omega}}}^m)\\
&+(2c\beta_{\mathrm{P}}\,{\bm{\mathrm{\tau}}}\wedge {\bm{\mathrm{e}}}^i+\gamma\varepsilon^i_{\phantom{i}jk}{\bm{\mathrm{e}}}^j\wedge {\bm{\mathrm{e}}}^k)\wedge\Big({\bm{\mathrm{d}}}{\bm{\mathrm{A}}}_i+\frac{1}{2}\varepsilon_{i\ell m}{\bm{\mathrm{A}}}^\ell\wedge {\bm{\mathrm{A}}}^m-\frac{c^2}{2}\varepsilon_{i\ell m}{\bm{\mathrm{\Omega}}}^\ell\wedge {\bm{\mathrm{\Omega}}}^m\Big)\,.
\end{align*}
When $c\to 0$, this Lagrangian becomes
\begin{equation}\label{HK_Carroll}
\mathcal{L}_0:=\gamma\varepsilon^i_{\phantom{i}jk}{\bm{\mathrm{e}}}^j\wedge {\bm{\mathrm{e}}}^k\wedge\Big({\bm{\mathrm{d}}}{\bm{\mathrm{A}}}_i+\frac{1}{2}\varepsilon_{i\ell m}{\bm{\mathrm{A}}}^\ell\wedge {\bm{\mathrm{A}}}^m\Big)\,.
\end{equation}
If, instead, we take $\gamma=0$, reabsorb the $c$ factor into the new constant $\kappa=c\beta_{\mathrm{P}}$, and take the limit $c\rightarrow 0$, we find
\begin{equation}\label{Action_Carroll_2}
\mathcal{L}_1=\kappa\left(2{\bm{\mathrm{\tau}}}\wedge {\bm{\mathrm{e}}}^i\wedge\Big({\bm{\mathrm{d}}}{\bm{\mathrm{A}}}_i\!+\!\frac{1}{2}\varepsilon_{i\ell m}{\bm{\mathrm{A}}}^\ell\wedge {\bm{\mathrm{A}}}^m\Big)\!+\!\varepsilon_{ijk}{\bm{\mathrm{e}}}^i\wedge {\bm{\mathrm{e}}}^j\wedge({\bm{\mathrm{d}}}{\bm{\mathrm{\Omega}}}^k\!+\!\varepsilon^{k}_{\phantom{k}\ell m}{\bm{\mathrm{A}}}^\ell\wedge {\bm{\mathrm{\Omega}}}^m)\right)\,.
\end{equation}
These two Lagrangians appear in the first two rows of Table 1 of \cite{Figueroa-OFarrill:2022mcy}. The first corresponds to the Husain-Kucha\v{r} model that we analyze in the following from the perspective of its Carrollian symmetry.

\section{Relevant facts about the Husain-Kucha\v{r} model}\label{sec_HK}

The action for the Husain-Kucha\v{r} (HK) model \cite{Husain:1990vz} is
\begin{equation}\label{HK_action}
  S(\mathbf{e},\mathbf{A}) := \frac{1}{2} \int_{\mathcal{M}} \varepsilon_{ijk} \mathbf{e}^i \wedge \mathbf{e}^j \wedge \mathbf{F}^k\,.
\end{equation}
Here $\mathcal{M}=\mathbb{R}\times\Sigma$ where $\Sigma$ is a closed (i.e. compact with no boundary) and orientable 3-dimensional manifold. Notice that this automatically implies that $\Sigma$ is parallelizable. The fields that appear in the action are 1-forms
\begin{align*}
{\bm{\mathrm{A}}}^i&\in\Omega^1(\mathcal{M})\,,&&i=1,2,3\,,\\
{\bm{\mathrm{e}}}^i&\in\Omega^1(\mathcal{M})\,,&&i=1,2,3\,.
\end{align*}
The curvature $\mathbf{F}^i$ is defined as
\[
{\bm{\mathrm{F}}}_i:=\bd{\bm{\mathrm{A}}}_i+\frac{1}{2}\varepsilon_{ijk}{\bm{\mathrm{A}}}^j\wedge{\bm{\mathrm{A}}}^k\,.
\]
and the covariant exterior differential acts on ${\bm{\mathrm{e}}}^i$ as
\[
{\bm{\mathrm{D}}}{\bm{\mathrm{e}}}_i:=\bd{\bm{\mathrm{e}}}_i+\varepsilon_{ijk}{\bm{\mathrm{A}}}^j\wedge {\bm{\mathrm{e}}}^k\,.
\]
$\mathcal{M}=\mathbb{R}\times\Sigma$ is foliated by $\Sigma_\tau:=\{\tau\}\times\Sigma$ for $\tau\in\mathbb{R}$ and we have a canonical vector field $\partial_t\in \mathfrak{X}(\mathcal{M})$, transverse to all the sheets $\Sigma_\tau$ of the foliation, defined by the tangent vectors to the curves $c_p:\mathbb{R}\rightarrow \mathcal{M}:\tau\mapsto (\tau,p)$ with $p\in\Sigma$. We also have a scalar field ${\bm t}\in C^\infty(\mathcal{M})$ defined as ${\bm t}:\mathcal{M}\rightarrow \mathbb{R}:(\tau,p)\mapsto \tau$. The vector field $\partial_t$ and the scalar field ${\bm t}$ satisfy ${\partial_t}\righthalfcup \bd \bm{t}=1$. For every $\tau\in\mathbb{R}$, we define the embedding $\jmath_\tau:\Sigma\rightarrow\mathcal{M}:p\mapsto(\tau,p)$ with its pullback denoted as $\jmath_\tau^\ast$. Notice that $\Sigma_\tau=\jmath_\tau(\Sigma)$.

The fact that $\Sigma$ is parallelizable allows us to restrict ourselves to field configurations such that
\[
\frac{1}{3!}\varepsilon_{ijk}\bd {\bm t}\wedge{\bm{\mathrm{e}}}^i\wedge{\bm{\mathrm{e}}}^j\wedge{\bm{\mathrm{e}}}^k=:\mathsf{vol}_{\bm t}
\]
is a volume form (and, as a consequence, the ${\bm{\mathrm{e}}}^i$ are linearly independent).

The field equations for \eqref{HK_action} are
\begin{align}\label{field_equations_HK}
\begin{split}
   & \varepsilon_{ijk}\bm{\mathrm{e}}^i\wedge {\bm{\mathrm{D}}} \bm{\mathrm{e}}^j=0\,, \\
   & \varepsilon_{ijk}\bm{\mathrm{e}}^j\wedge \bm{\mathrm{F}}^k=0\,.
   \end{split}
\end{align}

An alternative and useful perspective on the dynamics can be obtained from the Hamiltonian formulation for the HK model that we summarize now. The phase space $\Gamma$ is spanned by the objects $(e_{\mathrm{t}}^i,\widetilde{E}^i,A_{\mathrm{t}}^i,A^i)$ where, for each $i=1,2,3$ we have $e_{\mathrm{t}}^i,A_{\mathrm{t}}^i\in\Omega^0(\Sigma)$, $\widetilde{E}^i\in \mathfrak{X}(\Sigma)$, and $A^i\in\Omega^1(\Sigma)$. A vector field $\mathbb{Y}$ in this phase space has components $\big(Y_{e_{\mathrm{t}}}^i, Y_{\widetilde{E}}^i, Y_{A_{\mathrm{t}}}^i,Y_A^i\big)$. These components must be understood as maps
\begin{align*}
  Y_{e_{\mathrm{t}}}^i:\Gamma\rightarrow \Omega^0(\Sigma)\,, & \quad Y_{\widetilde{E}}^i:\Gamma\rightarrow \mathfrak{X}(\Sigma)\,, \\
  Y_{A_{\mathrm{t}}}^i\!:\Gamma\rightarrow \Omega^0(\Sigma)\,, & \quad Y_A^i:\Gamma\rightarrow \Omega^1(\Sigma)\,,
\end{align*}
for each $i=1,2,3$.

The relevant pre-symplectic form can be written formally as
\[
\omega=\int_\Sigma \left(\dd A^i\ww\dd \widetilde{E}_i\right) \mathsf{vol}_0\,,
\]
here $\mathsf{vol}_0$ is a fiducial, non-dynamical volume form on $\Sigma$ and $\dd$ and $\w$ denote the exterior differential and exterior product in $\Gamma$. The precise meaning of the previous expression can be understood by taking a pair of vector fields $\mathbb{X},\mathbb{Y}$ in phase space, and letting $\omega$ act on them \cite{BarberoG:2024fug}:
\begin{equation}\label{presymplectic}
\omega(\mathbb{X},\mathbb{Y})=\int_\Sigma \left(Y_{\widetilde{E}i}\righthalfcup  X_A^i-X_{\widetilde{E}i}\righthalfcup  Y_A^i\right)\mathsf{vol}_0\,.
\end{equation}
Notice that the integrand of this expression is a 3-form (as it should) and also that $\omega$ is degenerate because the $e_{\mathrm{t}}^i$ and $A_{\mathrm{t}}^i$ components of $\mathbb{X}$ and $\mathbb{Y}$ do not appear in it.

In addition to the symplectic form, we have constraints and Hamiltonian vector fields. The constraints are the familiar Gauss law and vector constraint of the Ashtekar formulation for General Relativity
\begin{align}
  & \mathrm{div}_0\widetilde{E}_i+\varepsilon_{ijk}{\widetilde{E}^k}\!\!\righthalfcup  A^j=0\,,\label{const_st_Ei1} \\
  & {\widetilde{E}_i}\righthalfcup  F^i=0\,.\label{const_st_Ei2}
\end{align}
Here $\mathrm{div}_0$ is the divergence defined by the fiducial volume form $\mathsf{vol}_0$.

The Hamiltonian vector fields are
\begin{align}
  & Z_A^i =\pounds_\xi A^i+D\Lambda^i\,, \label{vec_Z_A}\\
  & Z_{\widetilde{E}}^i =\pounds_\xi \widetilde{E}^i+(\mathrm{div}_0\xi) \widetilde{E}^i-\varepsilon^{ijk}\Lambda_j\widetilde{E}_k\,,\label{vec_Z_E}\\
  & Z_{A_{\mathrm{t}}}^i\quad \mathrm{arbitrary}\,,\label{vec_Z_Ai}\\
  & Z_{e_{\mathrm{t}}}^i\,\quad \mathrm{arbitrary}\,.\label{vec_Z_Ei}
\end{align}
In the previous expressions, $\Lambda^j(e_{\mathrm{t}}^i,\widetilde{E}^i,A_{\mathrm{t}}^i,A^i)=(A^j_{\mathrm{t}}-\xi\righthalfcup  A^j)$ and $\xi$ is a shorthand for
\begin{equation}\label{xi}
\xi(e_{\mathrm{t}}^i,\widetilde{E}^i)=\frac{e_{\mathrm{t}}^i\widetilde{E}_i }{\mathrm{det}\,e}\,, \quad {\mathrm{det}\,e}:=\sqrt{-\frac{1}{3!}\varepsilon_{ijk}\widetilde{E}^i\!\righthalfcup  \widetilde{E}^j\!\righthalfcup  \widetilde{E}^k \!\righthalfcup \mathsf{vol}_0}\,.
\end{equation}
Notice that $\pounds_\xi \widetilde{E}^i$ denotes the expression for the Lie derivative of $\widetilde{E}^i$ along $\xi(e_{\mathrm{t}}^i,\widetilde{E}^i)$ considering $\widetilde{E}^i$ as a vector field (not a vector density!). The curvature $F_i$ is
\[
F_i:=\mathrm{d}A_i+\frac{1}{2}\varepsilon_i^{\phantom{i}jk}A_j\wedge A_k\,,
\]
and the covariant exterior differential $D$ acts on the cotriads $e_i$ as
\[
De_i:=\mathrm{d}e_i+\varepsilon_i^{\phantom{i}jk}A_j\wedge e_k\,,
\]
with the usual generalization to differential forms of other valence.

Several comments are in order now. First, notice that the evolution of $A^i$ and $\widetilde{E}^i$ described by the Hamiltonian vector fields corresponds to a combination of spatial diffeomorphisms and internal $SO(3)$ transformations with parameters $(\Lambda,\xi)$ which, in turn, are ultimately defined in terms of $(e_{\mathrm{t}}^i,\widetilde{E}^i,A_{\mathrm{t}}^i,A^i)$.

Second, the fact that there are no conditions on $Z_{A_{\mathrm{t}}}^i$ and $Z_{e_{\mathrm{t}}}^i$ means that they are arbitrary smooth functions in phase space and, hence, $(e_{\mathrm{t}}^i,A_{\mathrm{t}}^i)$ are arbitrary in the sense that their evolution is governed by the equations
\begin{align*}
  \dot{e}_{\mathrm{t}}^i & =Z_{e_{\mathrm{t}}}^i\,, \\
  \dot{A}_{\mathrm{t}}^i & =Z_{A_{\mathrm{t}}}^i\,,
\end{align*}
which contain arbitrary elements. At this point, it is not obvious that we can just pick any function of time for $e_{\mathrm{t}}^i$ and $A_{\mathrm{t}}^i$ as our Lagrangian is time-independent (and, thus, the evolution equations do not depend explicitly on time). The right framework to do this would be one appropriate for time-dependent Lagrangians (which is not what we are using here). In any case, this arbitrariness is compatible with the fact that $e_{\mathrm{t}}^i$ and $A_{\mathrm{t}}^i$ do not show up in the presymplectic form $\omega$ given by \eqref{presymplectic}.

Finally, there is the interesting option of \textit{gauge fixing}. A useful way to carry it out consists of choosing a vector field $\hat{\xi}\in\mathfrak{X}(\Sigma)$ and imposing the gauge fixing conditions
\begin{align}
&\xi(e_{\mathrm{t}}^i,\widetilde{E}^i)=\hat{\xi}\,,\label{GF1}\\
&A_{\mathrm{t}}^i=\hat{\xi}\righthalfcup  A^i\,.\label{GF2}
\end{align}
By doing this, we immediately find
\begin{align}
&Z_A^i=\pounds_{\hat{\xi}}A^i\,,\label{vec_ZA_GF}\\
&Z_{\widetilde{E}}^i=\pounds_{\hat{\xi}}\widetilde{E}^i+(\mathrm{div}_0\hat{\xi}) \widetilde{E}^i \,.\label{vec_ZE_GF}
\end{align}
To complete the analysis, we have to figure out what the gauge fixing conditions tell us about $Z_{A_{\mathrm{t}}}^i$ and $Z_{e_{\mathrm{t}}}^i$ (and, thus, on the evolution of $A_{\mathrm{t}}^i$ and $e_{\mathrm{t}}^i$). First, from \eqref{GF2} we have
\[
Z_{A_{\mathrm{t}}}^i=\hat{\xi}\righthalfcup  Z_A^i=\hat{\xi}\righthalfcup \pounds_{\hat{\xi}}A^i=\pounds_{\hat{\xi}}(\hat{\xi}\righthalfcup  A^i)=\pounds_{\hat{\xi}}A_{\mathrm{t}}^i\,.
\]
Since $\hat{\xi}$ is defined by \eqref{xi}, we can rewrite \eqref{GF1} as
\[
e_{\mathrm{t}}^i=\hat{\xi}\righthalfcup  e^i\,,\quad e^i:=-\frac{1}{3!(\mathrm{det}\,e)}\varepsilon^i_{\phantom{i}jk}\widetilde{E}^j\!\righthalfcup \widetilde{E}^k \!\!\righthalfcup \mathsf{vol}_0\,,
\]
Notice that $\{e^i\}$ is, up to a $\mathrm{det}\,e$ factor, the dual basis of $\{\widetilde{E}^i\}$. Finally, we compute
\[
Z_{e_{\mathrm{t}}}^i=\hat{\xi}\righthalfcup \pounds_{\hat{\xi}}e^i=\pounds_{\hat{\xi}}(\hat{\xi}\righthalfcup  e^i)=\pounds_{\hat{\xi}}e_{\mathrm{t}}^i\,.
\]
As we can see, after performing the gauge fixing, the evolution of the fields $A_{\mathrm{t}}^i$ and $e_{\mathrm{t}}^i$ is no longer arbitrary but is also obtained by Lie-dragging them along $\hat{\xi}$. In conclusion, in this particular gauge, the dynamics of the HK model simply reduces to spatial diffeomorphisms generated by the vector field $\hat{\xi}$ used in the gauge fixing.

We end this section by pointing out that nothing precludes us from taking $\hat{\xi}=0$, in which case, we have no evolution at all: once we pick a solution to the constraints on an initial 3-surface $\Sigma_0$ and take $A_{\mathrm{t}}^i=0$ and $e_{\mathrm{t}}^i=0$ (as dictated by the gauge fixing conditions), the solution remains frozen. Moreover, to get a solution to the field equations \eqref{field_equations_HK} on $\mathcal{M}$, we can simply Lie drag these objects along $\partial_t$ (constant in $\tau$).

\section{Carrollian light cones and the dynamics of the Husain-Kucha\v{r} model}\label{sec_light_cones}

An interesting perspective on the dynamics given by \eqref{field_equations_HK} can be gained by introducing \cite{Husain:1990vz,BarberoG:2024fug}
\begin{equation}\label{def_u}
\tilde{u}_0(\cdot):=\frac{1}{3!}\left(\frac{\cdot\wedge \varepsilon_{ijk}\bm{\mathrm{e}}^i\wedge \bm{\mathrm{e}}^j\wedge\bm{\mathrm{e}}^k}{\mathsf{vol}_t }\right)\,,
\end{equation}
which can be interpreted as a vector field on $\mathcal{M}$ (that we also write as $\tilde{u}_0\in\mathfrak{X}(\mathcal{M})$, see the discussion in \cite{BarberoG:2024fug}). Here we denote the scalar field that multiplied by $\mathsf{vol}_t$ gives a certain top-form $\alpha$ on $\mathcal{M}$ as
\[
\left(\frac{\alpha}{\mathsf{vol}_t}\right)\,.
\]
Notice that, for a 1-form $\bm{\beta}\in\Omega^1(\mathcal{M})$, we have
\[
\bm{\beta}(\tilde{u}_0)=\tilde{u}_0(\bm{\beta})=\frac{1}{3!}\left(\frac{\bm{\beta}\wedge \varepsilon_{ijk}\bm{\mathrm{e}}^i\wedge \bm{\mathrm{e}}^j\wedge\bm{\mathrm{e}}^k}{\mathsf{vol}_t }\right)\in C^\infty(\mathcal{M})\,.
\]
A couple of interesting consequences of this are: $\tilde{u}_0\righthalfcup \bm{\mathrm{e}}^i=\bm{\mathrm{e}}^i(\tilde{u}_0)=0$ for $i=1,2,3$, and $\tilde{u}_0\righthalfcup {\bm{\mathrm{d}}}\bm{t}={\bm{\mathrm{d}}}\bm{t}(\tilde{u}_0)=1$. In practice, we are considering the non-degenerate coframe $({\bm{\mathrm{d}}}\bm{t},\bm{\mathrm{e}}^i)$ so that the element of the dual frame transverse to the foliation defined by $\bm{t}$ is $\tilde{u}_0$. Although in the following arguments we are relying on the foliation of $\mathcal{M}$ defined by $\bm{t}$ this is not strictly necessary: we can use any non-degenerate cotetrad built by completing the $\bm{\mathrm{e}}^i$ with an extra field $\bm{\tau}\in\Omega(\mathcal{M})$. Notice that the distribution defined by $\bm{\tau}$ need not be integrable. 

As discussed in \cite{Husain:1990vz} (see also \cite{BarberoG:2024fug}) if we take solutions $\bm{\mathrm{e}}^i$ and $\bm{\mathrm{A}}^i$ to the field equations \eqref{field_equations_HK}, and build the vector field $\tilde{u}_0$ with them, the following two conditions hold
\[
\tilde{u}_0\righthalfcup  {\bm{\mathrm{D}}} \bm{\mathrm{e}}^i=0\,,\quad \tilde{u}_0\righthalfcup \bm{\mathrm{F}}^i=0\,,
\]
and, as a consequence,
\begin{align*}
  &\pounds_{\tilde{u}_0}\bm{\mathrm{e}}^i=\varepsilon^{ijk}{\bm{\mathrm{e}}}_j(\tilde{u}_0\righthalfcup {\bm{\mathrm{A}}}_k)\,,\\
  &\pounds_{\tilde{u}_0}\bm{\mathrm{A}}^{\!\!i}\!={\bm{\mathrm{D}}}(\tilde{u}_0\righthalfcup {\bm{\mathrm{A}}}^i)\,.
\end{align*}
This means that Lie-dragging the solutions to the field equations along the integral curves defined by the vector field $\tilde{u}_0$ is equivalent to performing internal $SO(3)$ rotations with arbitrary infinitesimal parameter $\Upsilon^i=\tilde{u}_0\righthalfcup {\bm{\mathrm{A}}}^i$. As we see, there is a dynamically determined congruence of curves (the integral curves of $\tilde{u}_0$) that play a central role. The evolution along them is trivial in the sense that it will not affect physical observables. This neatly conforms with the picture of the Carrollian limit as that in which the light cones collapse to lines and the fact that ``light does not feel the passage of time''.

The previous approach to the dynamics of the model can also be understood from the Hamiltonian point of view presented above. Indeed, it is always possible to write
\[
\tilde{u}_0=\partial_t-\bm{\xi}
\]
for some $\bm{\xi}\in\mathfrak{X}(\mathcal{M})$ satisfying $\bm{\mathrm{d}}\bm{t}(\bm{\xi})=0$, i.e. $\bm{\xi}$ tangent to each $\Sigma_\tau$ for all $\tau\in\mathbb{R}$. If we define now $\bm{\mathrm{e}}^i_{\mathrm{t}}:=\partial_t\righthalfcup  \bm{\mathrm{e}}^i$ and $\bm{\mathrm{A}}_{\mathrm{t}}^i:=\partial_t\righthalfcup  \bm{\mathrm{A}}^i$ [both of them in $\Omega^0(\mathcal{M})$], we have
\begin{align*}
\pounds_{\tilde{u}_0}{\bm{\mathrm{A}}}^i&=\pounds_{\partial_t}{\bm{\mathrm{A}}}^i-\pounds_{\bm{\xi}}{\bm{\mathrm{A}}}^i={\bm{\mathrm{D}}}\big((\partial_t-\bm{\xi})\righthalfcup  {\bm{\mathrm{A}}}^i\big)\,,\\
\pounds_{\tilde{u}_0}{\bm{\mathrm{e}}}^i&=\pounds_{\partial_t}{\bm{\mathrm{e}}}^i-\pounds_{\bm{\xi}}{\bm{\mathrm{e}}}^i=\varepsilon^i_{\phantom{i}jk}{\bm{\mathrm{e}}}^j\big((\partial_t-\bm{\xi})\righthalfcup  {\bm{\mathrm{A}}}^k\big)\,,
\end{align*}
which give
\begin{align}
 \pounds_{\partial_t}{\bm{\mathrm{A}}}^i&=\pounds_{\bm{\xi}}{\bm{\mathrm{A}}}^i+{\bm{\mathrm{D}}}(\bm{\mathrm{A}}_{\mathrm{t}}^i-{\bm{\xi}}\righthalfcup  {\bm{\mathrm{A}}}^i)\,,\label{vec_1}\\
\pounds_{\partial_t}{\bm{\mathrm{e}}}^i&=\pounds_{\bm{\xi}}{\bm{\mathrm{e}}}^i+\varepsilon^i_{\phantom{i}jk}{\bm{\mathrm{e}}}^j(\bm{\mathrm{A}}_{\mathrm{t}}^k-{\bm{\xi}}\righthalfcup  {\bm{\mathrm{A}}}^k)\,.\label{vec_2}
\end{align}
If we pullback these expressions onto $\Sigma$ with the help of $\jmath_\tau^*$, we find
\begin{align}
\dot{A}^i(\tau)&=\pounds_{\xi(\tau)}A^i(\tau)+\bar{D}\Lambda^i(\tau)\,,\label{pull_back_1}\\
\dot{e}^i(\tau)&=\pounds_{\xi(\tau)}e^i(\tau)+\varepsilon^i_{\phantom{i}jk}e^j(\tau)\Lambda^k(\tau)\,.\label{pull_back_2}
\end{align}
Here
\begin{align*}
&\Lambda^i(\tau):=A_{\mathrm{t}}^i(\tau)-\xi(\tau)\righthalfcup  A^i(\tau)\,,\\
&  A_{\mathrm{t}}^i(\tau):=\jmath_t^\ast \bm{\mathrm{A}}_{\mathrm{t}}^i\,, && A^i(\tau):=\jmath_t^\ast \Aa^i\,,\\
&  \,e_{\mathrm{t}}^i(\tau)\,:=\jmath_t^\ast \bm{\mathrm{e}}_{\mathrm{t}}^i\,, && \,e^i(\tau)\,:=\jmath_t^\ast \bm{\mathrm{e}}^i\,,
\end{align*}
$\xi(\tau)$ is the restriction of $\bm{\xi}$ to $\Sigma_\tau$ (which is well-defined because $\bm{\xi}$ is tangent to each $\Sigma_\tau$), $\bar{D}$ is the covariant exterior differential defined by $A^i(\tau)$ and
\[
\dot{A}^i(\tau):=\jmath_t^\ast\pounds_{\partial_t}A^i\,,\quad \dot{e}^i(\tau):=\jmath_t^\ast\pounds_{\partial_t}e^i\,.
\]
If, with the help of the fiducial volume form $\mathsf{vol}_0$, we introduce the vector field 
\[
\widetilde{E}^i(\tau):=\frac{1}{2}\left(\frac{\cdot\wedge \varepsilon^i_{\phantom{i}jk}e^j(\tau)\wedge e^k(\tau)}{\mathsf{vol}_0}\right)
\]
and use \eqref{pull_back_2}, we find
\begin{equation}\label{pull_back_3}
\dot{\widetilde{E}} ^i(\tau)=\pounds_{\xi(\tau)}\widetilde{E}^i(\tau)+\big(\mathrm{div}_0\xi(\tau)\big)\widetilde{E}^i(\tau)-\varepsilon^i_{\phantom{i}jk}\widetilde{E}^j(\tau)\Lambda^k(\tau)\,.
\end{equation}
From \eqref{pull_back_1} and \eqref{pull_back_3}, we can read the form of the vector fields whose integral curves correspond to the solutions to the field equations. These are \textit{exactly} \eqref{vec_Z_A} and \eqref{vec_Z_E}.

\section{Conclusions and comments}\label{sec_conclusions}

We have analyzed in detail the Husain-Kucha\v{r} model from the perspective of its Carrollian symmetry. To this end, we have looked at the Carrollian limit of the Holst action, and discussed the dynamics of one of the resulting Lagrangians (the Husain-Kucha\v{r} model), both from a 4-dimensional and a Hamiltonian perspective. These two approaches provide different points of view. On one hand, by studying the field equations, it is possible to identify a congruence of curves which play a central role in the dynamical description of the system \cite{Husain:1990vz}. As the evolution of the dynamical field along these curves reduces to gauge transformations, physical observables remain constant on them. This is precisely what one expects in the Carrollian limit in which light cones collapse. It is also interesting to point out that the evolution along the curves tangent to the collapsed light cones effectively freezes.

The Hamiltonian perspective is an attractive alternative to understand the theory and its local symmetries, as they are expected to manifest themselves in the form of the Hamiltonian vector fields. As is well known, the dynamics of the HK model reduces to diffeomorphisms and local internal rotations. We have shown that, although this does not immediately lead to the interpretation of the dynamics mentioned in the preceding paragraph (which can be studied just by looking at the field equations), both points of view are actually compatible. In any case, by performing a suitable gauge fixing, the triviality of the evolution and its Carrollian features can also be seen in a satisfactory way in this setting. The analysis shows that the full Carrollian symmetry does not manifest itself at the Hamiltonian level, in particular, there is no trace of the Carrollian boosts. One reason for this is the absence  in the HK model of the fields $\bm{\tau}$ and $\bm{\Omega}^i$ (defined in Section \ref{defcarr}). At the end of the day one only sees diffeomorphisms and internal rotations, which are expected as they are manifest symmetries of the HK action. 

The HK model can be extended to manifolds with boundary where Chern-Simons theories play an interesting role. This may help understand boundary charges and the possible role of the collapsed Carrollian light cones. As the HK model is a simplified version of General Relativity it may provide a way to understand the symmetries of asymptotically flat spacetimes and related BMS charges on the boundary as suitable Carrollian limits (see \cite{Miskovic:2023zfz,Ecker:2024czh}). As a final remark, we want to point out that, in order to understand how the full Carrollian symmetry works one should consider the other Lagrangian \eqref{Action_Carroll_2} obtained in the Carrollian limit of the Holst action. If we are correct about how the Carrollian symmetry works, it should then be possible to find a congruence of curves similar to the one described for the HK model and such that the evolution of the fields in the direction of these curves reduces to gauge transformations. We will look at this problem in the near future.

\begin{appendices}

%
%

\section{Lie algebra contractions}\label{app_contractions}

Here we give some details on Lie algebra contractions that are relevant for the paper. A real Lie algebra $\mathfrak{g}=(V,[\cdot,\cdot])$ is a $\mathbb{R}$-vector space $V$ endowed with a bilinear, antisymmetric form $[\cdot,\cdot]$ satisfying the Jacobi identity
\begin{equation}\label{Jacobi}
[X,[Y,Z]]+[Y,[Z,X]]+[Z,[X,Y]]=0\,.
\end{equation}
In the following we will restrict ourselves to finite-dimensional vector spaces of dimension $n\in\mathbb{N}$. By chosing a linear basis $(b_i)$ on $V$ one can represent the bilinear form  $[\cdot,\cdot]$ by the structure constants $X^i_{\phantom{i}jk}\in (\mathbb{R}^n)^3$ according to
\[
[b_i,b_j]=X^k_{\phantom{k}ij}b_k\,,
\] 
with 
\[
X^i_{\phantom{i}jk}=-X^i_{\phantom{i}kj}\,,\quad X^l_{ij}X^n_{kl}+X^l_{jk}X^n_{il}+X^l_{ki}X^n_{jl}=0\,, 
\]
following from antisymmetry of the bracket and the Jacobi identity \eqref{Jacobi}. We can then think of all the possible a Lie algebra structures on the $n$-dimensional vector space $V$ as the ``Lie bracket submanifold'' $\mathcal{L}_n$ of $(\mathbb{R}^n)^3$ given by
\[
\mathcal{L}_n=\{X\in (\mathbb{R}^n)^3\,:\, X^i_{jk}+X^i_{kj}=0\,,\quad X^\ell_{ij}X^n_{k\ell}+X^\ell_{jk}X^n_{i\ell}+X^\ell_{ki}X^n_{j\ell}=0\}\subset (\mathbb{R}^n)^3\,.
\]
As $\mathcal{L}_n$ is defined as the vanishing set of some polynomials it should be thought of as an algebraic variety and not necessarily as a differentiable submanifold of $(\mathbb{R}^n)^3$. In order to define contractions of a Lie algebra, $\mathcal{L}_n$ must be endowed with the Zariski topology.

The general linear group $GL(V^n)$ acts on the left on $\mathcal{L}_n$ according to
\begin{align*}
GL(V^n)\times  \mathcal{L}_n\rightarrow \mathcal{L}_n\,,\quad    (g,X)\mapsto &g\cdot X\,, \quad(g\cdot X)^i_{\phantom{i}jk}:=g^i\,_{i'} (g^{-1})^{j'}\,_j(g^{-1})^{k'}\,_k X^{i'}_{\phantom{i'}j'k'}\,.
\end{align*}
The orbits of this action 
\[
\mathcal{O}(X)=\{g\cdot X\,:\, g\in GL(V^n)\}
\]
define the isomorphy classes of $n$-dimensional Lie algebras.

Given $X', X\in \mathcal{L}_n$, we say that $X$ degenerates to $X'$ ($X\rightarrow_{\mathrm{deg}} X'$) if $X'$ belongs to the closure of $X$ in the Zariski topology, $X'\in \overline{\mathcal{O}(X)}^Z$. A degeneration is said to be trivial if $X'\in \mathcal{O}(X)$ and non-trivial if $X'\in \partial^Z\mathcal{O}(X)$. In the case of real vector spaces, any degeneration is a \textit{contraction} \cite{Weimar-Woods} defined in the following way: Given a continuous $g:(0,1]\rightarrow GL(V^n)\,,\, \epsilon\mapsto g_\epsilon$ and fixing $X\in \mathcal{L}_n$, let us consider $g_\epsilon \cdot X$. If $\lim_{\epsilon\rightarrow 0} g_\epsilon \cdot X =X'$ exists and $X'\in \mathcal{L}_n$ we say that $X'$ is a contraction of $X$ (hence, a degeneration of $X$). For all $1\geq \epsilon>0$ it is evident that $g_\epsilon\cdot X$ defines an algebra isomorphic to $X$. As a consequence, if we want $X'$ not to be isomorphic to $X$, it is necessary (but not sufficient) that $\lim_{\epsilon\rightarrow 0} \mathrm{det}(g_\epsilon)=0$.

Homogeneous spaces furnish interesting examples of contractions. For instance, let us suppose that $\mathfrak{g}$ is a direct sum of vector spaces $\mathfrak{g}=\mathfrak{h}\oplus\mathfrak{m}$; $[\mathfrak{h},\mathfrak{h}]\subset \mathfrak{h}$ ($\mathfrak{h}$ is a Lie subalgebra of $\mathfrak{g}$); $[\mathfrak{h},\mathfrak{m}]\subset \mathfrak{m}$ ($\mathrm{ad}:\mathfrak{h}\times \mathfrak{m}\rightarrow \mathfrak{m}\,, (h,m)\mapsto \mathrm{ad}_hm=[h,m]$ is an action) and, finally, $[\mathfrak{m},\mathfrak{m}]\neq\{0\}$. We take now a basis $b=(b_{\mathfrak{h}},b_{\mathfrak{m}})$ of $\mathfrak{g}$ and rescale the elements of $b_{\mathfrak{m}}$ as $(b_\mathfrak{h},b_\mathfrak{m})\mapsto (b'_\mathfrak{h},b'_\mathfrak{m})=(b_\mathfrak{h},\epsilon\, b_\mathfrak{m})$ with $\epsilon>0$. By doing this we get
\[
\begin{array}{l}
{[b_\mathfrak{h},b_\mathfrak{h}]= b_\mathfrak{h}}\\
{[b_\mathfrak{h},b_\mathfrak{m}]= b_\mathfrak{m}}\\
{[b_\mathfrak{m},b_\mathfrak{m}] =b_\mathfrak{h}+b_\mathfrak{m}}
\end{array}\rightsquigarrow\quad
\begin{array}{l}
{[b_\mathfrak{h},b_\mathfrak{h}] = b_\mathfrak{h}}\\
{[b_\mathfrak{h},b'_\mathfrak{m}]= b'_\mathfrak{m}}\\
{[b'_\mathfrak{m},b'_\mathfrak{m}]= \epsilon^2 (b_\mathfrak{h}+\epsilon^{-1}b'_\mathfrak{m})= \epsilon^2 b_\mathfrak{h}+\epsilon b'_\mathfrak{m}}
\end{array}
\]
and in the limit $\epsilon\rightarrow 0$ we find
\[
\begin{array}{l}
{[b_\mathfrak{h},b_\mathfrak{h}] = b_\mathfrak{h}}\\
{[b_\mathfrak{h},b'_\mathfrak{m}]= b'_\mathfrak{m}}\\
{[b'_\mathfrak{m},b'_\mathfrak{m}]= 0}
\end{array}\rightsquigarrow
\begin{array}{l}
{[\mathfrak{h},\mathfrak{h}]\subset\mathfrak{h}}\\
{[\mathfrak{h},\mathfrak{m}]\subset\mathfrak{m}}\\
{[\mathfrak{m},\mathfrak{m}]=0}
\end{array}
\]
The contracted algebra $\mathfrak{g}_c$ is a semidirect product of Lie algebras: 
\begin{align*}
[(h_1,m_1),(h_2,m_2)]&:=[h_1\oplus m_1,h_2\oplus m_2]=[h_1,h_2]\oplus([h_1,m_2]+[m_1,h_2])\\
&\,\;=([h_1,h_2],\mathrm{ad}_{h_1} m_2-\mathrm{ad}_{h_2}m_1)
\end{align*}
and exponentiating we get the Lie group $G_c=H\ltimes M$. 

We discuss now some examples. Some classic results are $\mathfrak{so}(p+1,q)\rightarrow_{\mathrm{deg}} \mathfrak{iso}(p,q)$ and $\mathfrak{so}(p,q+1)\rightarrow_{\mathrm{deg}} \mathfrak{iso}(p,q)$, 
in particular  $\mathfrak{so}(p+1)\rightarrow_{\mathrm{deg}} \mathfrak{iso}(p)$. In order to see the latter let us take $\mathfrak{so}(p+1)=\mathrm{span}(L_{ab})$, where the $L_{ab}$ satisfy
\[
 [L_{ab},L_{cd}]=\eta_{ad}L_{cb}-\eta_{ac}L_{db}-\eta_{bd} L_{ca}+\eta_{bc}L_{da}\,.
\]
with $\eta_{ab}=\mathrm{diag}(-1_{p+1})$. Let us consider also $\mathfrak{iso}(p)=\mathrm{span}(L_{ab},P_c)=\mathbb{R}^p\rtimes \mathfrak{so}(p)$ with 
\begin{align*}
    &\hspace*{1.9cm}[L_{ab},L_{cd}]&\hspace*{-1.8cm}&=\eta_{ad}L_{cb}-\eta_{ac}L_{db}-\eta_{bd} L_{ca}+\eta_{bc}L_{da}\,,\\
    &\hspace*{1.9cm}[L_{ab},P_c]&\hspace*{-1.8cm}&=\eta_{ca}P_b-\eta_{cb}P_a\,,\\
    &\hspace*{1.9cm}[P_a,P_b]&\hspace*{-1.8cm}&=0\,.
\end{align*}
We split now the $L_{ab}$ in $\mathfrak{so}(p+1)$ as 
\[
L_i:=L_{i,p+1}\,,\quad L_{ij}\,,\quad 1\leq i,j\leq p
\]
in such a way that $\mathfrak{h}=\mathrm{span}(L_{ij})$ and $\mathfrak{m}=\mathrm{span}(L_i)$. The structure constants can be read from
\begin{align*}
&[L_{ij},L_{kl}]=\delta_{il}L_{kj}-\delta_{ik}L_{lj}-\delta_{jl} L_{ki}+\delta_{jk}L_{li}\,,\\
&[L_{ij},L_k]=\delta_{ki}L_j-\delta_{kj}L_i\,,\\
&[L_i,L_j]=-L_{ij}\,.
\end{align*}
Notice that $[\mathfrak{h},\mathfrak{h}]\subset \mathfrak{h}$ and $[\mathfrak{h},\mathfrak{m}]\subset \mathfrak{m}$. We define now $P_i =\epsilon L_i$ with $\epsilon\in(0,1]$ and get
\begin{align*}
&[L_{ij},L_{kl}]=\delta_{il}L_{kj}-\delta_{ik}L_{lj}-\delta_{jl} L_{ki}+\delta_{jk}L_{li}\,, \\
&[L_{ij},P_k]=\delta_{ki}P_j-\delta_{kj}P_i\,, \\
&[P_i,P_j]=-\epsilon P_{ij}\,,
\end{align*}
which in the limit $\epsilon\rightarrow0$ gives
\begin{align*}
& [L_{ij},L_{kl}]=\delta_{il}L_{kj}-\delta_{ik}L_{lj}-\delta_{jl} L_{ki}+\delta_{jk}L_{li}\,,\\
&[L_{ij},P_k]=\delta_{ki}P_j-\delta_{kj}P_i\,,\\
&[P_i,P_j]=0\,,
\end{align*}
i.e. the $\mathfrak{iso}(p)$ Lie algebra. 

\end{appendices}

%
%

\end{document}